\documentclass[iop]{emulateapj}

\usepackage{color}
\usepackage{float}
\usepackage{epstopdf}
\usepackage{graphicx}
\usepackage[utf8]{inputenc}
\usepackage{natbib}

\shortauthors{Martis et al.}
\begin{document}
\title{The Evolution of the Fractions of Quiescent and Star-forming Galaxies as a Function of Stellar Mass Since z=3: Increasing Importance of Massive, Dusty Star-forming Galaxies in the Early Universe}

\author{Nicholas S. Martis$^{1*}$, Danilo Marchesini$^{1}$, Gabriel B. Brammer$^{2}$, Adam Muzzin$^{3}$, Ivo Labb\'e$^{4}$, Ivelina G. Momcheva$^{2}$, Rosalind E. Skelton$^{5}$, Mauro Stefanon$^{4}$, Pieter G. van Dokkum$^{6}$, \and Katherine E. Whitaker$^{7 \dagger}$ }
\footnotetext[*]{nicholas.martis@tufts.edu}
\footnotetext[$\dagger$]{Hubble Fellow}
\affil{$^1$Department of Physics and Astronomy, Tufts University, Medford, MA 02155, USA\\
$^2$Space Telescope Science Institute, 3700 San Martin Drive, Baltimore,
MD 21218, USA\\
$^3$Kavli Institute for Cosmology, University of Cambridge, Cambridge, CB3 0HA, United Kingdom\\
$^4$Leiden Observatory, Leiden University, PO Box 9513, NL-2300 RA Leiden,
The Netherlands\\
$^5$South African Astronomical Observatory PO Box 9, Observatory, Cape Town, 7935, South Africa\\
$^6$Department of Astronomy, Yale University, 260 Whitney Avenue, New Haven, CT 06511, USA\\
$^7$Department of Astronomy, University of Massachusetts, Amherst, MA 01003, USA }

\begin{abstract}
Using the UltraVISTA DR1 and 3D-HST catalogs, we construct a stellar-mass-complete sample, unique for its combination of surveyed volume and depth, to study the evolution of the fractions of quiescent galaxies, moderately unobscured star-forming galaxies, and dusty star-forming galaxies as a function of stellar mass over the redshift interval $0.2 \le z \le 3.0$. We show that the role of dusty star-forming galaxies within the overall galaxy population becomes more important with increasing stellar mass, and grows rapidly with increasing redshift. Specifically, dusty star-forming galaxies dominate the galaxy population with $\log{(M_{\rm star}/M_{\odot})} \gtrsim 10.3$ at $z\gtrsim2$. The ratio of dusty and non-dusty star-forming galaxies as a function of stellar mass changes little with redshift. Dusty star-forming galaxies dominate the star-forming population at $\log{(M_{\rm star}/M_{\odot})} \gtrsim 10.0-10.5$, being a factor of $\sim$3-5 more common, while unobscured star-forming galaxies dominate at $\log{(M_{\rm star}/M_{\odot})} \lesssim 10$. At $\log{(M_{\rm star}/M_{\odot})} > 10.5$, red galaxies dominate the galaxy population at all redshift $z<3$, either because they are quiescent (at late times) or dusty star-forming (in the early universe).

\end{abstract}


\section{Introduction}
In the last decade, studies of the evolution  of the stellar mass function of galaxies have provided relatively robust measurements of the buildup of stellar mass in the Universe over most of cosmic history (Madau \& Dickinson 2014 for a review). It has been shown that the number density of quiescent galaxies grows dramatically with cosmic time as star-forming galaxies quench \citep[e.g.,][]{bram11, muz13b, ilbert13, tom14}. Additionally, in contrast to the hierarchical growth of dark matter halos predicted by N-body simulations, the most massive galaxies (i.e., $\log{(M_{\rm star}/M_{\odot})} \ge 11$) have been found to assemble their stellar mass and quench earlier than less massive galaxies \citep[e.g.,][]{pg08, mar09, mar10, font09, muz13b}. 

Driven by the realization that baryonic processes are the critical ingredients to reproduce the observed reversal of the hierarchical growth of structures, the latest theoretical models of galaxy formation match the observed evolution of the stellar mass function of galaxies in the last 11.5 Gyr of cosmic history reasonably well \citep[e.g.,][]{guo11, hen13, hen15, vog14, crain15}.  Despite the recent success, the fraction of quiescent galaxies as a function of stellar mass and redshift has proven challenging to match \citep[e.g.,][]{hen13, hen15}. Complicating the task is the tight interplay among the many baryonic processes associated with star formation and quenching. A robust measurement of the evolution of the fraction of quiescent galaxies as a function of stellar mass is therefore a tremendously powerful tool for making progress in our understanding of the feedback processes responsible for quenching and their relevant timescales. 

Recent near-infrared (NIR) studies have extended the measurements of the fraction of quiescent galaxies from intermediate redshifts \citep[e.g.,][]{bram11, ilbert13, muz13b} all the way to $z\sim4$ \citep[for the most massive galaxies; e.g.,][]{mar10, spitler14, str14}. Observationally, there are two main challenges in measuring robust fractions of quiescent galaxies. First, both wide and deep NIR surveys are required to probe the galaxy population at the high- and low-mass ends, as well as to progressively higher redshifts. Second, the selection of quiescent galaxies based on rest-frame colors, as mostly performed by previous works, can potentially be affected by the effects of dust obscuration and reddening. Dusty starburst galaxies can contribute a significant fraction to quiescent samples selected on the basis of a single rest-frame color (e.g., $U-V$) \citep{bram09}. Evidence for increasing dust obscuration with increasing stellar mass has been brought forward by \citet{whit12}, who have previously shown that massive dusty galaxies comprise a growing fraction of the galaxy population out to $z \sim 2$ \citep{whit10}.

Indeed, the population of dusty star-forming (massive) galaxies in the early universe appears to be the typical progenitors of today's most massive galaxies \citep{mar14}, and their link to both the overall star-forming population and the quiescent galaxies across time can provide crucial clues in furthering our understanding of galaxy evolution.

With this Letter we aim to provide a complete census of quiescent as well as both mildly obscured and dusty star-forming galaxies by measuring the evolution of their fractions as a function of redshift and stellar mass using the unique combination of the UltraVISTA DR1 and 3D-HST datasets.

This Letter is organized as follows: In section 2 we present the data. Section 3 describes the analysis. In section 4 we present our results and discuss our conclusions. All magnitudes are in the AB system. We assume a cosmology with $\Omega_\Lambda$ = 0.7, $\Omega_M$ = 0.3, and $H_0$ = 70 km s$^{-1}$Mpc$^{-1}$. 

\begin{figure*}[ht]
\includegraphics[width=\textwidth]{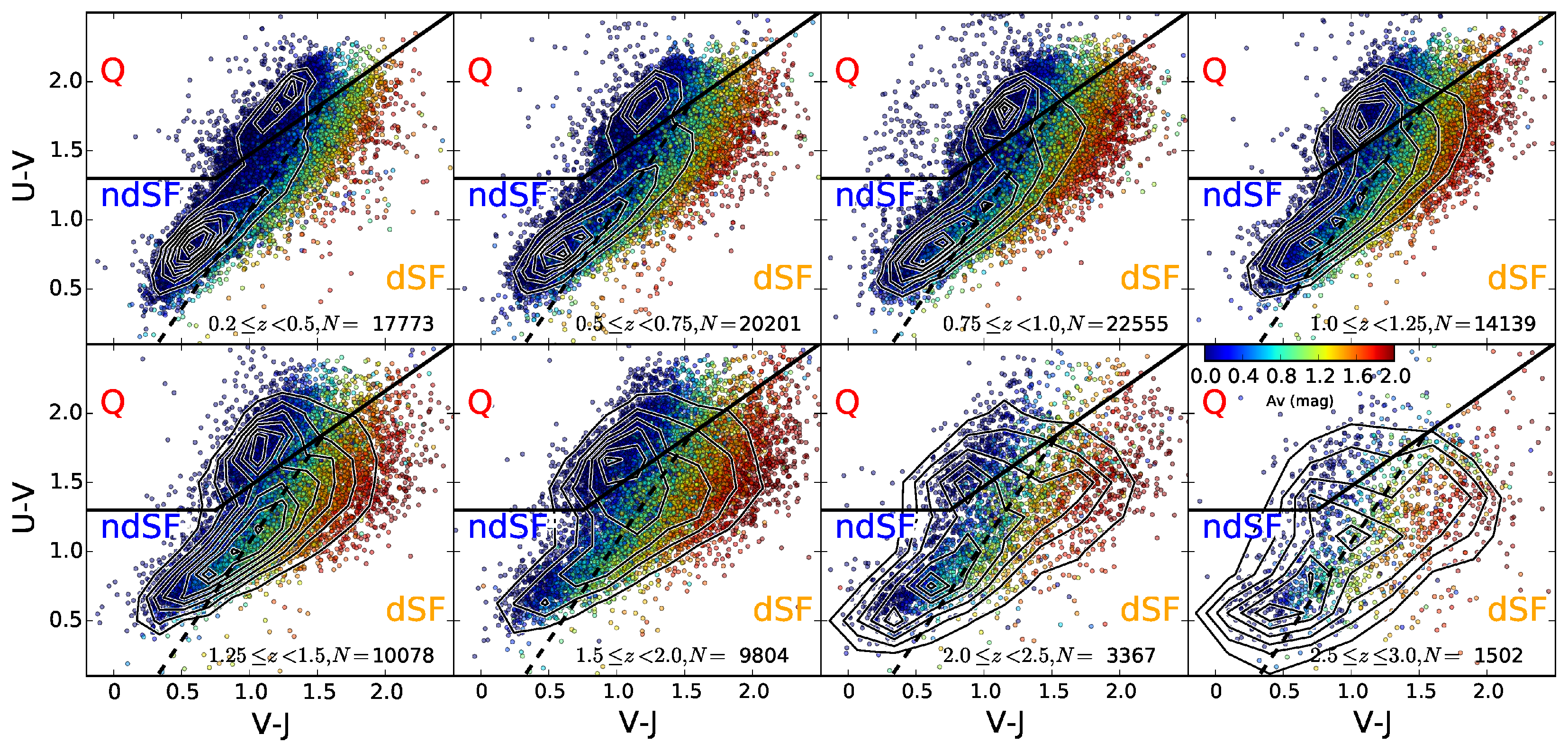}
\caption{UVJ diagram for the combined UltraVISTA and 3D-HST sample; galaxies are color coded as a function of $A_{\rm V}$ as determined using FAST. Each panel corresponds to a different redshift range. The number of sources in each redshift bin is shown. The contour curves indicate the density of sources. The three regions in the UVJ diagram to classify quiescent, non-dusty star-forming, and dusty star-forming galaxies are highlighted by Q (red), ndSF (blue), and dSF (orange), respectively.  \label{fig1}}
\end{figure*}

\section{Data and Sample}

We used the UltraVISTA DR1 \citep{muz13a} and the 3D-HST \citep{skel14, mom15} catalogs. 

The UltraVISTA v4.1 dataset is a $K_S$-selected photometric catalog covering 1.62 deg$^2$ with photometry in 30 bands. The catalog is 90\% complete to a depth of $K_{S,tot} = 23.4$. Unlike the public catalogs, we derived photometric redshifts and rest-frame colors adopting an old and dusty template \citep{mar10, bram16} in addition to the EAZY \citep{bram09} templates used in Muzzin et al. (2013a). \citet{muz13a} showed that including this template reduces the number density of $\log{(M_{\rm star}/M_{\odot})} > 11$ galaxies above $z \sim 2.5$ by $0.2-0.4$ dex. Stellar masses are determined using FAST \citep{kriek09} to fit the galaxy SEDs using \citet{bc03} stellar population synthesis models, a \citet{chab03} IMF, an exponentially declining star formation history, a \citet{cal00} extinction curve, and solar metallicity. The 95\% completeness in stellar mass as a function of redshift is taken from \citet{muz13a}, which contains a more detailed description of the catalog construction.

The 3D-HST WFC3-selected photometric v4.1 catalogs cover ~900 arcmin$^2$ over five fields. G141 grism redshifts from 3D-HST for 22,548 of the detected galaxies down to JH=24 (10.8\% of the photometric catalog) were provided by \citet{mom15}. For each galaxy, we use the best available redshift, namely spectroscopic, grism, and photometric. We refer to \citet{bez16} for an assessment of the quality of the grism and photometric redshifts of 3D-HST. Stellar masses were derived with FAST using the same SED-modeling assumptions as for UltraVISTA. We use the 90\% completeness level which corresponds to $H_{F160W} \le 25.1$ \citep{skel14}. The survey reaches 90\% completeness down to $\log{(M_{\rm star}/M_{\odot})} = 8.5$ and $\log{(M_{\rm star}/M_{\odot})} = 10$ at $z \le 0.5$ and $z \le 3$, respectively. 

Since we use the rest-frame $U-V$ and $V-J$ colors to define quiescent and star-forming galaxies, it is critical that the rest-frame colors of the UltraVISTA and 3D-HST survey are homogeneously derived. We therefore exploited the overlapping regions within the COSMOS field to compare the rest-frame colors. First, we matched sources in COSMOS from the UltraVISTA  and 3D-HST surveys within a separation of 0.2". Second, we quantified the offsets in rest-frame $U-V$ and $V-J$ as a function of redshift between the UltraVISTA and 3D-HST COSMOS sources. Third, the offsets as a function of redshift were fitted with a polynomial function, and the resulting best-fit model was used to adjust the UltraVISTA colors, homogenizing the two surveys. We note that the offsets in the rest-frame colors reach a maximum of $\sim 0.07$ mag in $U-V$ and $\sim 0.1$ mag in $V-J$. Though small, these offsets need to be corrected to ensure consistent classification of galaxies in the two surveys. The small difference in colors between the two catalogs is most likely the result of different zeropoint offsets adopted in each dataset and is thus not unexpected. Most importantly, we verified that, after the aforementioned homogenization, quantitatively similar results are obtained using, separately, the overlapping catalogs of the UltraVISTA and 3D-HST surveys. 

The combination of the UltraVISTA DR1 and 3D-HST results in a unique sample that allows us 1) to minimize cosmic variance at both the high- and low-mass ends; 2) to probe down to $M_{\rm star} \approx 10^{10}$~M$_{\odot}$ over the entire studied redshift range, and down to $M_{\rm star} \approx 10^{9}$~M$_{\odot}$ at $z\sim0.6$; 3) to sample the high-mass end up to $\log{(M_{\rm star}/M_{\odot})} \sim 11.5$ with good statistics; and 4) to quantify and mitigate systematic uncertainties between the two surveys using the overlapping region in the COSMOS field between UltraVISTA and 3D-HST. The final sample contains 99,419 galaxies at $0.2 \le z \le 3$ after removing the UltraVISTA objects falling in the patch of sky surveyed by 3D-HST in the COSMOS field. Spectroscopic and grism redshifts are available for 7,393 and 9,265 galaxies, respectively.

\section{Analysis}
For each dataset, the sample is divided into non-dusty star-forming, dusty star-forming, and quiescent galaxy populations using locations in the rest frame $U-V$ and $V-J$ color-color (UVJ, hereafter) diagram. Using the UVJ diagram to separate star-forming and quiescent galaxies has become a standard tool in the field \citep[e.g.][]{wuyts07, will, forrest}. Following \citet{whit15}\footnote[1]{$V-J < 1.5$ is no longer implemented, as it is a false upper limit imposed on the quiescent population \citep{vanderwel14}.}, star-forming galaxies satisfy: 

\begin{equation}
(U-V) <  1.3 \textrm{ for }(V-J) < 0.75
\end{equation}
and
\begin{equation}
(U-V) <  0.8(V-J) + 0.7 \textrm{ for } (V-J) \ge 0.75
\end{equation}

We introduced an additional criterion to further separate the star-forming population into dusty and relatively unobscured star-forming galaxies. Specifically, dusty star-forming galaxies satisfy:
\begin{equation}
(U-V) < 1.43(V-J) - 0.36.
\end{equation}

This division between dusty and unobscured star-forming galaxies was determined empirically using two independent approaches to estimate the dust obscuration in the rest-frame visual band, $A_{\rm V}$. In the first approach, we used $A_{\rm V}$ as determined from FAST. Due to the reddening degeneracy between age and obscuration, the derived $A_{\rm V}$ may be dependent on our assumption of an exponentially declining star-formation history. The second approach, in which $A_{\rm V}$ was determined from EAZY using the relative contribution to the observed SED of the dusty templates, does not suffer from this degeneracy in the same way. Figure 1 shows the stellar mass complete sample in the UVJ diagrams at the eight redshift intervals targeted in this work. The galaxies are color coded as a function of $A_{\rm V}$ as derived from FAST. The separation between quiescent and star-forming galaxies is shown as a black solid line. The separation between relatively unobscured and dusty star-forming galaxies (dashed black line) was defined to trace the ridge separating star-forming galaxies with $A_{\rm V} < 1$ (non-dusty star-forming galaxies, ndSF, hereafter) and those with $A_{\rm V} > 1$ (dusty star-forming galaxies, dSF, hereafter). We note that Figure 1 does not change quantitatively if the $A_{\rm V}$ from EAZY is adopted in place of the $A_{\rm V}$ from FAST. This gives us confidence that our criterion to separate dSF and ndSF galaxies is robust. Figure 1 also shows contours representing the density of sources in the UVJ diagram.

\begin{figure*}[ht]
\includegraphics[width=\textwidth]{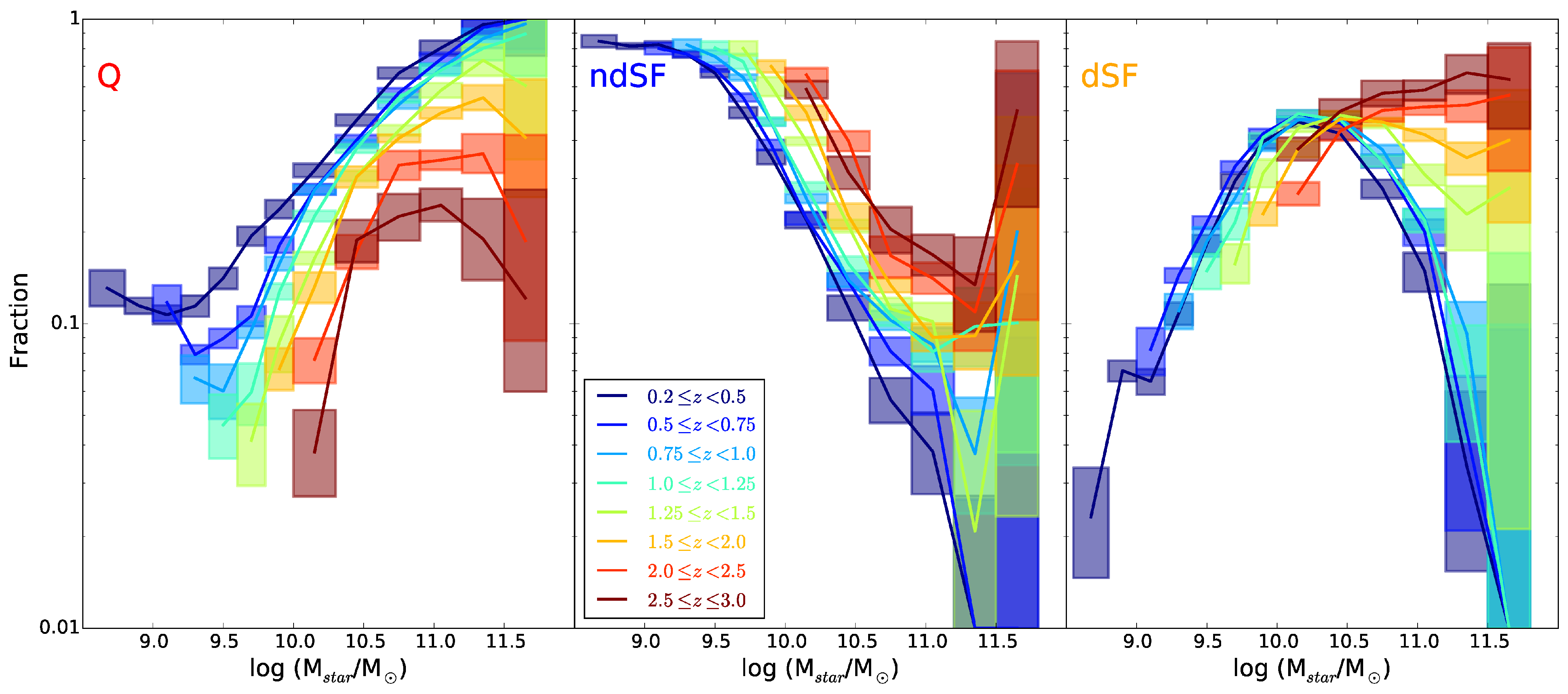}
\caption{The fraction of quiescent (left), ndSF (center), and dSF (right) galaxies as a function of stellar mass for eight bins in redshift as indicated by color. Shaded regions represent 1$\sigma$ total errors; the widths represent the size of the stellar mass bin. \label{fig2}}
\end{figure*}

\begin{figure*}[ht]
\includegraphics[width=\textwidth]{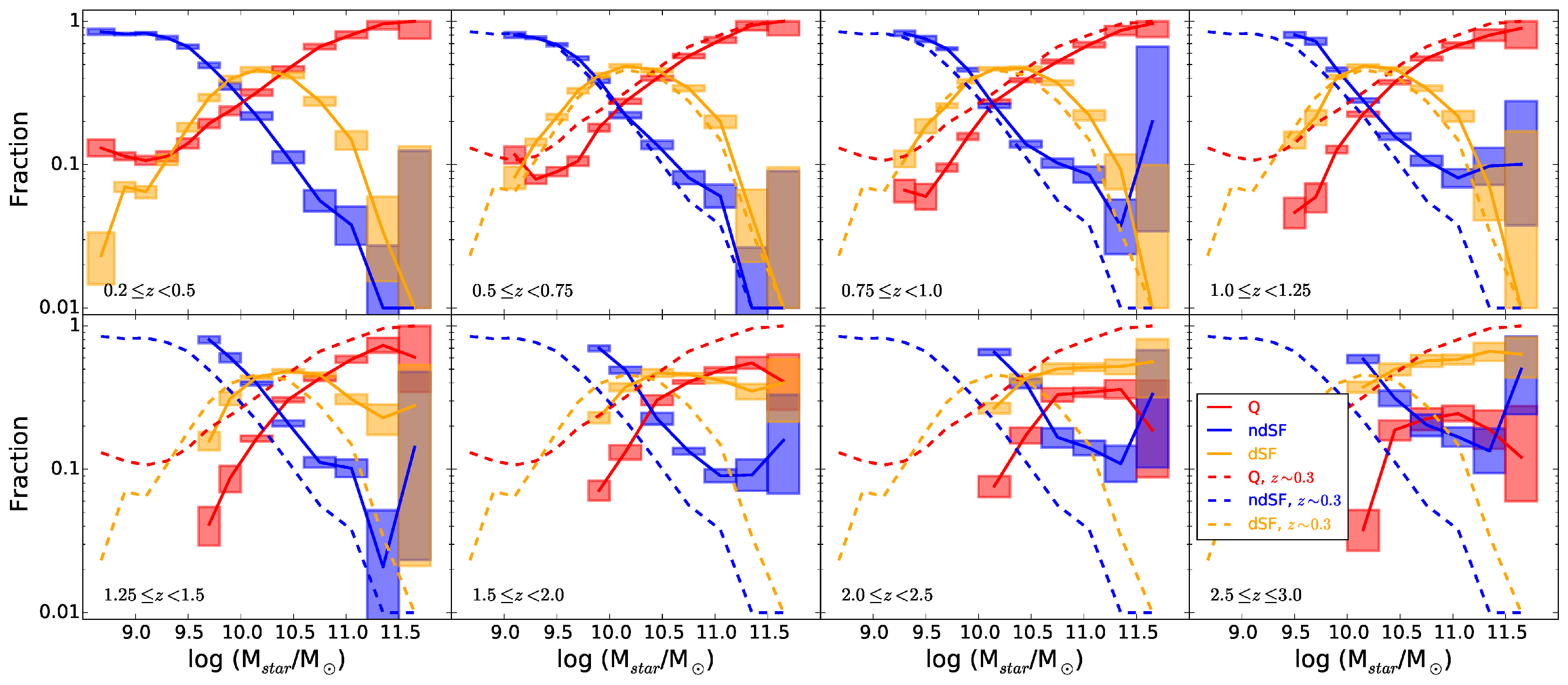}
\caption{The fraction of quiescent (red), ndSF (blue), and dSF (orange) galaxies as a function of stellar mass in each redshift bin. Shaded regions represent the 1$\sigma$ total errors. The fractions at $0.2 \le z < 0.5$ are overplotted as dashed curves to highlight the  evolution. \label{fig3}}
\end{figure*}

Figure 1 shows the strong bimodality at low redshift in the galaxy population, with quiescent and star-forming galaxies well separated in the UVJ diagram, as previously shown \citep{will, bram11, muz13b}. While the bimodality in the UVJ diagram is present out to $z \sim 2.5$, we note an increase in the relative number of dSF galaxies with increasing redshift. Specifically, at $z \ge 1.5$, we see the star-forming population progressively migrating toward the region occupied by highly dust-obscured star-forming galaxies, while simultaneously, the population of quiescent galaxies becomes less prominent.

In order to quantify the evolution seen in the UVJ diagram, we calculated the  fraction of quiescent, ndSF, and dSF galaxies as a function of stellar mass. The fractions, and associated errors, in each redshift and stellar mass bin were calculated separately for 3D-HST and UltraVISTA. The total error on the fractions includes Poisson errors (from \citealt{g}), as well as the error due to uncertainties in redshifts and stellar masses. The latter were estimated performing 200 Monte Carlo simulations of the catalogs. For each realization, we independently perturbed the photometric redshifts and the stellar masses using the 68\% confidence limits as derived by EAZY and FAST. The fractions of quiescent, ndSF, and dSF galaxies were then recalculated, allowing for the derivation of the errors on the fractions from the 1$\sigma$ scatter in the resulting distribution of fractions. Finally, the fraction of quiescent, ndSF, and dSF galaxies in each bin of stellar mass and redshift were derived as the average of the separately calculated fractions in UltraVISTA and 3D-HST weighted by their respective errors. 

\section{Results and Discussion}
Figure 2 shows the evolution with redshift of the fractions of quiescent, ndSF and dSF galaxies. Several results can be noted from Figure 2. Focusing first on the population of quiescent galaxies, we confirm the progressive building up of the quiescent population with cosmic time, as already found previously by, e.g., \citet{bram11} and \citet{muz13b}. We robustly see that the building up of the quiescent population proceeds fastest for the most massive galaxies (i.e., $\log{(M_{\rm star}/M_{\odot})}\approx11.5$) and slowest for galaxies with $\log{(M_{\rm star}/M_{\odot})}\approx10.5$. Interestingly, at $\log{(M_{\rm star}/M_{\odot})}\lesssim10.3$, the evolution with cosmic time of the fraction of quiescent galaxies appears to quicken, potentially indicative of the progressively increasing role of environmental effects,  arguably the dominant quenching mechanism at the low-mass end, at later times \citep[e.g.,][]{peng2010, peng2012, woo2013}. The fraction of quiescent galaxies plateaus at $\sim$10\% at $\log{(M_{\rm star}/M_{\odot})}<9.5$ at $z\sim0.35$.

Correspondingly, the fraction of star-forming galaxies decreases with increasing cosmic time. Focusing on the ndSF galaxies, Figure 2 shows that, at all redshifts, they represent the dominant galaxy type at the low-mass end, while their fraction rapidly decreases with increasing stellar mass. The corresponding stellar mass at which the ndSF galaxies dominate the overall population (i.e., fraction $>$50\%) is seen to decrease with cosmic time, evolving from $\sim2\times10^{10}$~M$_{\odot}$ at $z\sim2.75$ to $\sim5\times10^{9}$~M$_{\odot}$ at $z\sim0.35$. 

The right panel of Figure 2 shows that at $z<1.25$, dSF galaxies reach $\sim$50\% of the overall galaxy population at $10\lesssim \log{(M_{\rm star}/M_{\odot})}\lesssim10.5$, where their relative importance rapidly decreases at both lower and higher stellar masses. However, at $z>1.25$, dSF galaxies rapidly increase their importance at the high-mass end. Specifically, while dSF galaxies represent $\sim$60\% of the  population at $\log{(M_{\rm star}/M_{\odot})}\sim11.5$ at $z\sim2.75$, their fraction is seen to drop by a factor of $\sim30$ by $z\sim0.35$. At $z>2$, dSF galaxies constitute $\sim$50-60\% of the overall population at $\log{(M_{\rm star}/M_{\odot})}>10.5$. This shows that massive star-forming galaxies are predominantly heavily dust obscured in the early universe.

Figure 3 shows the comparison of the fractions of quiescent (red), ndSF (blue), and dSF (orange) at each redshift. Figure 3 better highlights which of the three populations dominates in a particular range in stellar mass and redshift. It is clear that at $z>2$, the dSF galaxies are the dominant population at $\log{(M_{\rm star}/M_{\odot})} \gtrsim 10.3$, whereas at $\log{(M_{\rm star}/M_{\odot})} \lesssim 9.8-10.3$ (with the smaller value at low redshift), ndSF galaxies are most common. At $z<1.5$, dSF galaxies are the dominant type of galaxies only in the stellar mass range $10.0 \lesssim \log{(M_{\rm star}/M_{\odot})} \lesssim 10.5$. Figure 3 clearly shows the growth of the quiescent galaxies with cosmic time at the expense of the star-forming galaxies, with the growth of the quiescent population happening first at the highest masses, and then shifting to lower stellar masses with decreasing redshift. Finally, Figure 3 shows that for $z<3$, the high-mass end (i.e., $\log{(M_{\rm star}/M_{\odot})} \gtrsim10.5$), is predominantly populated by red galaxies, either because they are quiescent (at late times) or dusty star-forming (in the early universe).

\begin{figure}
\includegraphics[width=\columnwidth]{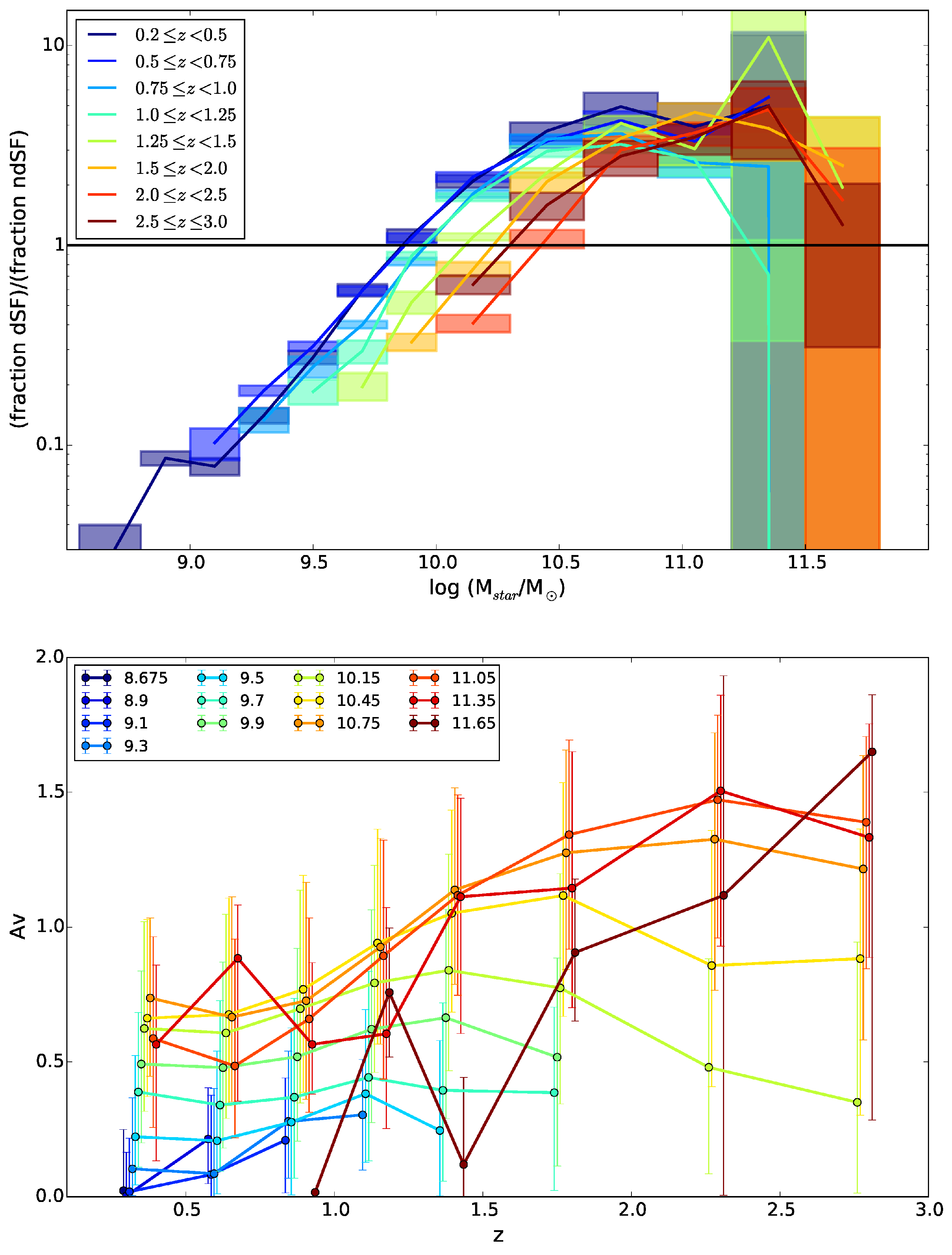}
\caption[width=\columnwidth]{Top: The ratio of the fraction of dSF galaxies to the fraction of ndSF galaxies as a function of stellar mass. Shaded regions represent total 1$\sigma$ errors. Bottom: The median $A_{\rm V}$ for all star-forming galaxies as a function of redshift in bins of stellar mass. Errorbars indicate the $25^{th}$ and $75^{th}$ percentiles. Points are offset slightly for clarity. \label{fig4}}
\end{figure}

An outstanding issue is the relative importance of dSF galaxies within the overall star-forming population as a function of stellar mass, and how their role may evolve with redshift. The top panel of Figure 4 shows the fraction of dSF galaxies divided by the fraction of ndSF galaxies as a function of stellar mass at each redshift. We see that dSF galaxies are a factor of $\sim$3-5$\times$ more important at the high-mass end ($\log{(M_{\rm star}/M_{\odot})}\gtrsim10.5$) than ndSF galaxies, which dominate at the low-mass end. The stellar mass dependency of the importance of dSF galaxies among the star-forming population appears to be very similar at all redshifts, monotonically increasing with stellar mass, with the dSF galaxies taking over at stellar masses $\log{(M_{\rm star}/M_{\odot})}\approx10.0-10.5$. These results are in qualitative agreement with \citet{whit12} and \citet{pan}, who observed a correlation between stellar mass and dust obscuration in star-forming galaxies. 

In order to better understand the previous findings, in the bottom panel of Figure 4 we show the median $A_{\rm V}$ as determined by FAST as a function of redshift in different stellar mass bins for all star-forming galaxies. Error bars indicate the 25$^{th}$ and 75$^{th}$ percentiles. First, we note that there is a large scatter for all stellar masses considered, and increasingly so at higher masses. At the low-mass end ($\log{(M_{\rm star}/M_{\odot})}\lesssim10.3$), we do not see much evolution with redshift in the amount of dust obscuration. At the high-mass end, however, the median $A_{\rm V}$ of star-forming galaxies increases rapidly with redshift. We therefore see that the population of massive star-forming galaxies becomes increasingly obscured with increasing redshift, despite the relative number of massive dSF and ndSF galaxies not significantly evolving with redshift (top panel of Figure 4).

It is tempting to compare the results shown in Figure 4 with the evolution of the stellar mass metallicity relation observed in star-forming galaxies from $z\sim3$ \citep[e.g.][]{maiolino08, sanders15, kashino16, wuyts16}, which proceeds faster for lower stellar masses. For example, \citet{maiolino08} inferred that since $z\sim3.5$, the average metallicity of galaxies with $M_{\rm star}\approx10^{11}$~M$_{\odot}$ and $M_{\rm star}\approx10^{10}$~M$_{\odot}$ increased by $\sim0.5$~dex and $\sim0.8$ dex respectively. At fixed redshift, more massive galaxies appear more metal rich than lower mass galaxies. If we naively assume that increased metallicity leads to increased obscuration, our results only agree with some of these findings, potentially due to selection effects.  Specifically, we see that at $M_{\rm star}\approx10^{10}$ M$_{\odot}$, the relative role of dSF galaxies increases with cosmic time, as expected from the observed metal enrichment with cosmic time. We also find that dSF galaxies tend to become the predominant type among the star-forming population at larger stellar masses at earlier times, although this is only marginally significant. The bottom panel of Figure 4 finally shows that at a fixed redshift, the median amount of obscuration increases with stellar mass, in qualitative agreement with the mass-metallicity relation. However, the observed trend of increasing obscuration with redshift for massive, star-forming galaxies appears inconsistent with the evolution of the mass-metallicity relation, whether the studied samples may be biased against dusty galaxies \citep[e.g.][]{kashino16}, or arguably more representative of the overall galaxy population \citep[e.g.][]{sanders15}. Our results suggest that measurements of the mass-metallicity relation may indeed be missing significant numbers of heavily obscured star-forming galaxies at high redshift. The exact nature of the coupling between metallicity and dust formation complicates this comparison, and more quantitative analysis, beyond the scope of this Letter, will be necessary to assess any potential inconsistencies. 

Our results are affected by significantly large uncertainties at the very massive end, i.e., $\log{(M_{\rm star}/M_{\odot})}\gtrsim11.5$. These galaxies are extremely rare, and even the UltraVISTA DR1 $\sim$1.7~deg$^{2}$ survey does not sample a large enough volume to find enough ultra-massive galaxies. Recently completed or on-going multi-wavelength NIR surveys (e.g., NMBS-II and VIDEO), imaging up to 12~deg$^{2}$, will enable the construction of significantly larger samples of distant ultra-massive galaxies. Finally, although the UVJ diagram is quite robust at separating quiescent, ndSF and dSF galaxies, more sophisticated techniques to estimate the dust obscuration and level of star-formation will be necessary to improve the presented analysis.

\section{acknowledgments}
DM acknowledges the support of the Research Corporation for Science Advancement’s Cottrell Scholarship. DM and NM acknowledge the National Science Foundation under Grant No. 1513473. Support from STScI grant GO-1277 is gratefully acknowledged. KEW gratefully acknowledges support by NASA through Hubble Fellowship grant\#HF2-51368 awarded by the Space Telescope Science Institute, which is operated by the Association of Universities for Research in Astronomy, Inc., for NASA. Based on data products from observations made with ESO Telescopes at the La Silla Paranal Observatory under ESO programme ID 179.A-2005 produced by TERAPIX and the Cambridge Astronomy Survey Unit on behalf of the UltraVISTA consortium. 


\begin{table*}[h]
\centering
\label{Table 1}
\caption{Fractions (and total 1$\sigma$ errors) of quiescent, ndSF, and dSF galaxies}
\resizebox{\textwidth}{!}{%
\begin{tabular}{llllll}
 \hline
 Redshift & Stellar Mass Bin  & $f_{\rm Q}$ & $f_{\rm ndSF}$ & $f_{\rm dSF}$ & Number\\ [0.5ex]
 \hline \\ [-1.8ex]
$0.20 \le z < 0.50$ & $8.55 < \log{(M_{\rm star}/M_{\odot})} \le 8.80$  & 0.131$_{-0.017}^{+0.019}$ & 0.846$_{-0.039}^{+0.040}$ & 0.023$_{-0.009}^{+0.010}$ & 1503 \\ [1ex]
... & $8.80 < \log{(M_{\rm star}/M_{\odot})} \le 9.00$  & $0.115_{-0.007}^{+0.007}$  & $0.815_{-0.017}^{+0.017}$ & $0.070_{-0.006}^{+0.006}$ & 3188 \\ [1ex]
... & $9.00 < \log{(M_{\rm star}/M_{\odot})} \le 9.20$  & $0.107_{-0.007}^{+0.008}$ & $0.825_{-0.018}^{+0.019}$ & $0.065_{-0.006}^{+0.006}$ & 2641 \\ [1ex]
... & $9.20 < \log{(M_{\rm star}/M_{\odot})} \le 9.40$  & $0.114_{-0.009}^{+0.010}$ & $0.767_{-0.020}^{+0.021}$ & $0.108_{-0.009}^{+0.009}$ & 2050 \\ [1ex]
... & $9.40 < \log{(M_{\rm star}/M_{\odot})} \le 9.60$  & $0.141_{-0.011}^{+0.012}$ & $0.664_{-0.021}^{+0.022}$ & $0.183_{-0.013}^{+0.013}$ & 1716 \\ [1ex]
... & $9.60 < \log{(M_{\rm star}/M_{\odot})} \le 9.80$  & $0.194_{-0.014}^{+0.014}$ & $0.493_{-0.020}^{+0.021}$ & $0.294_{-0.017}^{+0.017}$ & 1498 \\ [1ex]
... & $9.80 < \log{(M_{\rm star}/M_{\odot})} \le 10.00$ & $0.238_{-0.016}^{+0.017}$ & $0.353_{-0.019}^{+0.020}$ & $0.395_{-0.020}^{+0.020}$ & 1307 \\ [1ex]
... & $10.00 < \log{(M_{\rm star}/M_{\odot})} \le 10.30$ & $0.319_{-0.016}^{+0.017}$ & $0.219_{-0.013}^{+0.014}$ & $0.456_{-0.019}^{+0.019}$ & 1639 \\ [1ex]
... & $10.30 < \log{(M_{\rm star}/M_{\odot})} \le 10.60$ & $0.466_{-0.020}^{+0.021}$ & $0.113_{-0.010}^{+0.011}$  & $0.421_{-0.019}^{+0.019}$  & 1431 \\ [1ex]
... & $10.60 < \log{(M_{\rm star}/M_{\odot})} \le 10.90$ & $0.665_{-0.027}^{+0.029}$ & $0.056_{-0.009}^{+0.010}$ & $0.278_{-0.020}^{+0.020}$  & 995 \\ [1ex]
... & $10.90 < \log{(M_{\rm star}/M_{\odot})} \le 11.20$ & $0.804_{-0.043}^{+0.046}$ & $0.038_{-0.010}^{+0.013}$  & $0.149_{-0.022}^{+0.022}$ & 462  \\ [1ex]
... & $11.20 < \log{(M_{\rm star}/M_{\odot})} \le 11.50$ & $0.960_{-0.081}^{+0.040}$ & $0.007_{-0.007}^{+0.017}$ & $0.034_{-0.019}^{+0.026}$ & 151  \\ [1ex]
... & $11.50 \le \log{(M_{\rm star}/M_{\odot})} \le 11.80$ & $1.0_{-0.243}^{+0.0}$    & $0.0_{-0.0}^{+0.115}$ & $0.0_{-0.0}^{+0.125}$ & 17   \\ [1ex]
$0.50 \le z < 0.75$ & $9.00 < \log{(M_{\rm star}/M_{\odot})} \le 9.20$  & $0.117_{-0.015}^{+0.016}$ & $0.801_{-0.035}^{+0.036}$ & $0.082_{-0.014}^{+0.015}$ & 2275 \\ [1ex]
... & $9.20  < \log{(M_{\rm star}/M_{\odot})} \le 9.40$  & $0.079_{-0.006}^{+0.006}$ & $0.768_{-0.016}^{+0.016}$ & $0.144_{-0.008}^{+0.008}$ & 3364 \\ [1ex]
... & $9.40  < \log{(M_{\rm star}/M_{\odot})} \le 9.60$  & $0.089_{-0.006}^{+0.007}$ & $0.687_{-0.017}^{+0.017}$ & $0.215_{-0.010}^{+0.010}$ & 2913 \\ [1ex]
... & $9.60  < \log{(M_{\rm star}/M_{\odot})} \le 9.80$  & $0.106_{-0.008}^{+0.008}$  & $0.552_{-0.016}^{+0.017}$ & $0.328_{-0.013}^{+0.013}$  & 2508 \\ [1ex]
... & $9.80  < \log{(M_{\rm star}/M_{\odot})} \le 10.00$ & $0.181_{-0.011}^{+0.011}$ & $0.386_{-0.015}^{+0.016}$ & $0.420_{-0.016}^{+0.016}$ & 2166 \\ [1ex]
... & $10.00 < \log{(M_{\rm star}/M_{\odot})} \le 10.30$ & $0.276_{-0.011}^{+0.012}$ & $0.221_{-0.010}^{+0.011}$ & $0.488_{-0.015}^{+0.015}$ & 2870 \\ [1ex]
... & $10.30 < \log{(M_{\rm star}/M_{\odot})} \le 10.60$ & $0.401_{-0.015}^{+0.015}$  & $0.137_{-0.009}^{+0.009}$ & $0.456_{-0.016}^{+0.016}$ & 2279 \\ [1ex]
... & $10.60 < \log{(M_{\rm star}/M_{\odot})} \le 10.90$ & $0.572_{-0.020}^{+0.022}$ & $0.081_{-0.008}^{+0.009}$ & $0.342_{-0.017}^{+0.017}$ & 1592 \\ [1ex]
... & $10.90 < \log{(M_{\rm star}/M_{\odot})} \le 11.20$ & $0.739_{-0.034}^{+0.035}$ & $0.060_{-0.010}^{+0.012}$ & $0.200_{-0.020}^{+0.020}$ & 729  \\ [1ex]
... & $11.20 < \log{(M_{\rm star}/M_{\odot})} \le 11.50$ & $0.938_{-0.074}^{+0.062}$ & $0.008_{-0.008}^{+0.016}$ & $0.044_{-0.023}^{+0.023}$ & 175  \\ [1ex]
... & $11.50  < \log{(M_{\rm star}/M_{\odot})} \le 11.80$ & $1.0_{-0.202}^{+0.0}$    & $0.0_{-0.0}^{+0.080}$ & $0.0_{-0.0}^{+0.086}$ & 26  \\ [1ex]
\hline \\
\end{tabular}%
}
Complete version available online.
\end{table*}


\clearpage

\end{document}